\documentclass[aps,pre,showpacs,floatfix,twocolumn]{revtex4-1}

\usepackage{amsmath}
\usepackage{amsfonts}
\usepackage{amssymb}
\usepackage{graphics}
\usepackage{graphicx}  
\begin{document}
\title{Results on ``Three predictions for July 2012 Federal Elections in Mexico based on past
regularities''}


\author{H. Hern\'andez-Salda\~na}   

\affiliation{Departamento de Ciencias B\'asicas, Universidad Aut\'onoma Metropolitana at Azcapotzalco, D.F., Mexico}

\email{E-mail: hhs@correo.azc.uam.mx}

\begin{abstract}
The Presidential Election in Mexico of July 2012 has been the third time that  PREP, 
Previous Electoral Results Program works. PREP gives voting outcomes based in 
electoral certificates of each polling station that arrive to capture centers. 
In previous ones, some statistical regularities 
had been observed, three of them were selected to make predictions and were published 
in \texttt{arXiv:1207.0078 [physics.soc-ph]}. Using the database made public in July 2012, 
two of the predictions were completely fulfilled, while, the 
third one was   
measured and confirmed using the database obtained upon request to the electoral authorities.
The first two predictions confirmed by actual measures are: (ii) The Partido Revolucionario 
Institucional, PRI, is a sprinter and has a better performance in polling stations 
arriving late to capture centers during the process. (iii) Distribution of vote of this party is well described by a 
smooth function named a Daisy model. A Gamma distribution, but compatible with 
a Daisy model, fits the distribution as well. The third prediction confirms that {\it errare humanum est},
since the error distributions of all the self-consistency variables appeared as a central 
power law with lateral lobes as in 2000 and 2006 electoral processes.  The three measured 
regularities appeared no matter the political environment.
\end{abstract}

\maketitle

\section*{Introduction}
Even when the study and modeling of electoral statistics is an area of traditional interest for
Political Economy and, in general, Political Sciences, the availability of databases 
in the last two decades made electoral systems an area amenable to study for physicists
and mathematicians. A wide variety of theoretical models with this point of view exist (see for instance 
\cite{CastellanoReport} and references therein) and in the last decade the number of studies of  
actual (empirical) data is growing
\cite{Castellano2007,CostaFilho1999,CostaFilho2003698,CostaFilho2009,Herrmann2006,HernandezSaldanaE1,Borghesi2012a,Borghesi2012b,Fortunato2013}. 
Predictions on the 
analysis of such data are appearing together with some theoretical frameworks trying to 
explain the regularities found in the ``experimental'' data. Notice that these approaches 
are far from those made by traditional political scientists.

Between the predictions we remark are those of Borghesi \cite{Borghesi09a}
which have been verfied\cite{Borghesi09b}. Here we present the results for 
three predictions made before the July 2012 Mexican electoral process and made public in
\cite{HernandezSaldanaE4}. As we shall see, two of them were fulfilled 
and the third one was incomplete due to the change in the official data presentation, 
which forbade the publication of the self-consistency data  while the certificates were
processed.

\section*{Results and Discussion}

\subsection*{Data and Observables}
The analysis is performed on  the dataset provided by the electoral authorities through
the {\it Programa de Resultados Electorales Previos}, PREP or Previous Electoral 
Results Program, during the election day and the next one. On how this program is implemented see 
the official electoral authorities 
web page\cite{IFEweb,IFE_prep} or reference \cite{HernandezSaldanaE4}. On the peculiarities of the 
Mexican electoral processes seei, for instance, \cite{Klesner}. Upon request, the electoral authorities 
gave access to the self-consistency additional data and the corresponding analysis is 
presented\cite{infomex_ife}. 

The database for the whole process contains the fields recording polling stations IDs, number of 
votes for each political party/candidate, time of arrival and a set of control fields that are summarized in 
Table \ref{tabla2}. We consider $139,657$ valid records, from a total of $144,013$, in the dataset 
provided by the electoral authorities\cite{infomex_ife}. For 2000 and 2006 we used the dataset
from references \cite{IFEweb,IFE_prep}. 

For the first prediction we use the distribution (not a normalized histogram) of the variables E$_i$ which are built up from the 
values of the fields described in Table \ref{tabla2}; there, six independent combinations
available are considered. The variables are built in order to see the lack or excess of votes in the records, 
for instance the total number of voters must coincide with the number of deposited ballots in the urn (E$_4$). 
In the ideal case all the distributions must be  Dirac's delta functions. 
So, these distributions are, in fact, the error distributions.

\begin{table}[ht]
\caption{
\bf{Self consistency fields and errors considered in the PREP database of July 2000, 2006 and 2012.}}

\begin{tabular}{|c|c|c|}
\hline

E$_1$ & B. received                              & Br - (Bs + V) \\
      & $-$ (B. not used $+$ Number of voters)   &                \\ \hline
E$_2$ & B. received                              & Br - (Bs + Bd) \\
      &  $-$ (B. not used $+$ B. deposited)      &     \\  \hline
E$_3$ & B. received                              & Br - (Bs+$\sum_i$ V$_i$) \\
      &   $-$ (B. not used $+$ Votes for each party)    &      \\  \hline
E$_4$ & Number of voters                         & V - Bd  \\
      &        $-$  B. deposited                  &     \\ \hline
E$_5$ & Number of voters & V - $\sum_i$ V$_i$ \\
      & $-$ Votes for each party                 &              \\   \hline
E$_6$ & B. deposited            & Bd - $\sum_i$ V$_i$ \\
      & $-$ Votes for each party                 &              \\ 
\hline
\end{tabular}
\begin{flushleft}{
We abbreviated Ballots with B.. The variable V$i$ stands for the number of votes obtained
for each party/candidate.
}
\end{flushleft}
\label{tabla2}
\end{table}

For the second prediction 
we consider the percentage of votes
for a party (PRI, in the current study) at a certain time $t$ or at a certain percentage of processed votes
certificates.  For the third prediction instead of the percentage of votes obtained we 
consider the distribution of votes: it is made by the histogram of the number of polling 
stations with certain amount of votes, properly scaled and normalized
in order to have a distribution normalized to unity with unit mean as well. 
Since we wish to  compare with a probability 
distribution the amount of votes is ``unfolded'' or ``deconvoluted'' by using 
the average number of votes, which properly scales the variable. The resulting
histogram must be normalized to area one.
(In reference \cite{HernandezSaldanaE2} there is
an explanation, but this procedure is standard in data treatment.) 
For simplicity we focus only on results from the presidential
election.

\begin{figure}[!ht]
\begin{center}
\includegraphics[width=0.9\columnwidth,angle=0]{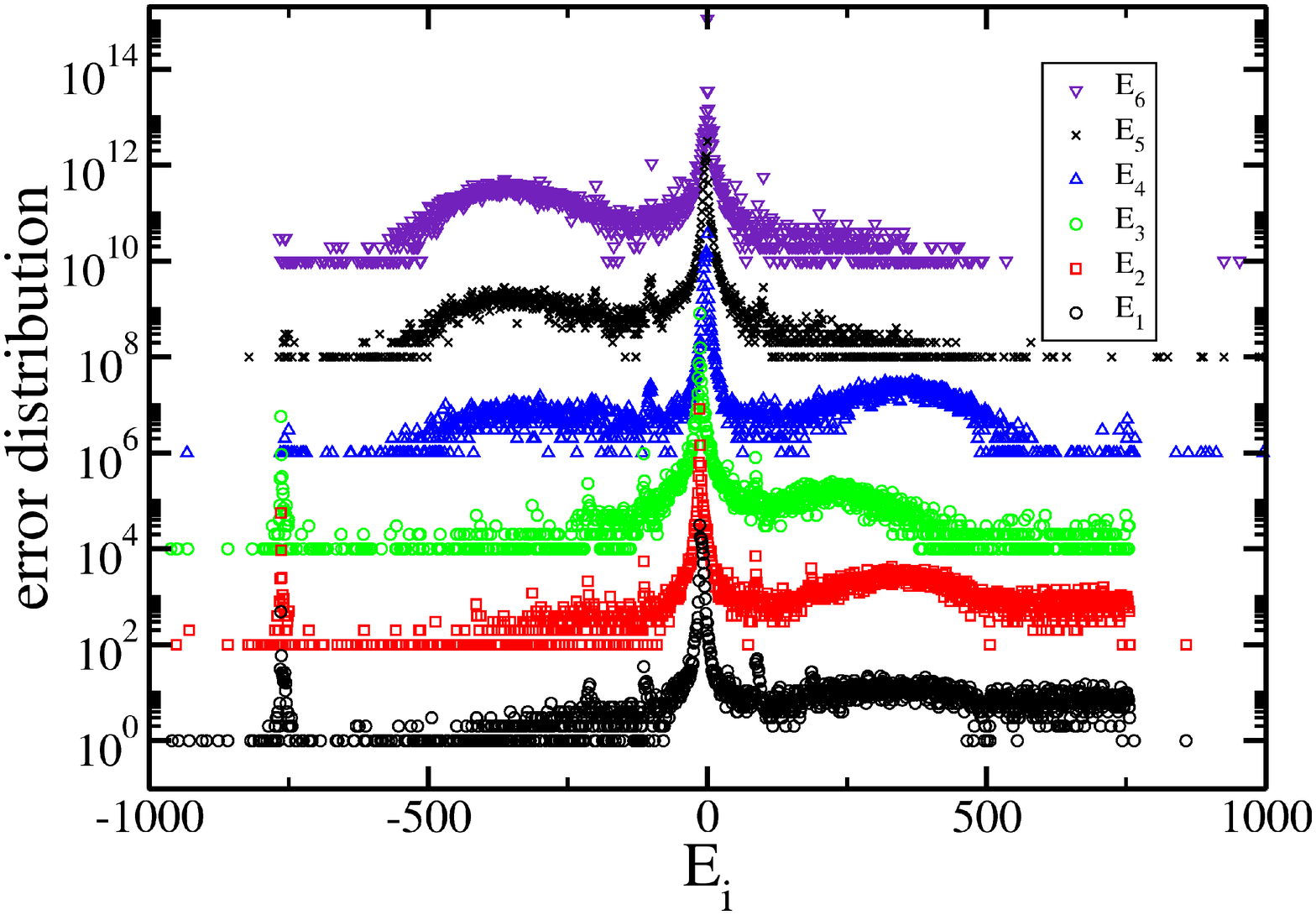}
\end{center}
\caption{{\bf Error distributions for the variables described 
in Table \ref{tabla2} (from bottom to top) for the presidential election on July 2012 in Mexico.}
The distribution are plotted in log-linear scale. We scale by a factor of 
$100$ each time.
Notice that they are characterized by a power law at the center and asymmetric lobes at each side.}
\label{fig:4}
\end{figure}

\begin{figure}[!ht]
\begin{center}
\includegraphics[width=0.9\columnwidth,angle=0]{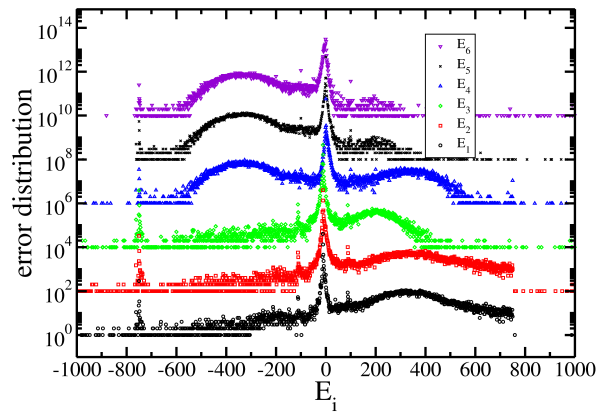}
\end{center}
\caption{{\bf Same as previous for 
the presidential election of July 2000 in Mexico.}
See text for details.} 
\label{fig:5}
\end{figure}

\subsection*{ Prediction  i) Errors could be epidemic in contemporary Mexican elections}
Self-consistency records in electoral data are an important measure in order to test and 
understand the sources of errors. The distribution of such records were presented in the 
first version of \cite{HernandezSaldanaE0} for the July 2006 federal 
elections presidential and for both chambers and the presidential election of 2000 in an ulterior version. The
regularities present for the error distributions described in Table \ref{tabla2},
allowed me to formulate the following prediction:

\bigskip
{\it Error distributions in self consistency tests of PREP's dataset will be
described globally by a power law at the center and two asymmetric lobes
at each side.}
\bigskip

\begin{figure}[ht]
\begin{center}
\includegraphics[width=0.9\columnwidth,angle=0]{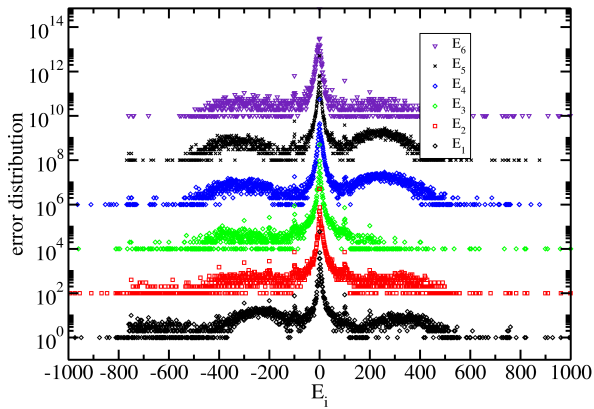}
\end{center}
\caption{{\bf  Same as previous for 
the mexican presidential election of July 2006.}
See text for details.} 
\label{fig:6}
\end{figure}

The six independent distributions are shown in Fig. \ref{fig:4} for the presidential process of 
July 2012.
As can be seen there the whole behaviour is similar to that in the distributions 
calculated for the 2000 (see Fig. \ref{fig:5}) and 
2006 (Fig. \ref{fig:6}) processes. 
As explained below they are the histograms of 
the number of cabins that have values of error equal to 0,1,2,$\cdots$.
Certainly, it can be interpreted as  
appearance and missing of votes, but in fact it is a measure on how we count the electoral 
results. The errors can be intentional cheating or counting mistakes. For the present case some changes appeared in the 
labels of the dataset, for instance, the ``Total number of Ballots deposited'', Bd, are now
``Total number of extracted ballots from the urn''; the sum of all the votes for the different
political parties and candidates is now made directly by  the IFE's  computers. 
In the last case, we test the value calculated by the computers and the sum on 
the records with no main differences. It is important to notice that the final 
electoral results are presented and accounted after all the parties reached an agreement on 
the results in each polling station.  

In all the figures the distributions are scaled by a factor of $100$ each time and in a
log-linear graph in order to appreciate all of them in a single figure. 
In all the cases the central part has a power law decay and two asymmetrical lobes. 
Large peaks appear in all the graphs in Figs. \ref{fig:4} to \ref{fig:6}. 
The reasons for this behaviour is unclear but it looks as a general feature 
that deserves a wide and detailed study\cite{HernandezSaldana_IIbarra}.

\subsection*{ Prediction ii) The Partido Revolucionario Institucional (PRI) is a sprinter}

Even when the behaviour presented here for the presidential candidates of the Partido 
Revolucionario Institucional appeared in election for the both chambers we shall concentrate in the 
presidential case. 
A graph 
of the percentage of votes for each party/candidate against the percentage of computed polling stations
had been presented in voters outcome reports for federal elections in 2000 and 2006. In reference \cite{HernandezSaldanaE0}, version 3, both elections are reported.
In Figure 1 and 2 of 2006, and in Figure 9 for 2000 the results are presented. In all the
analyzed cases, the PRI showed a change in the percentage of votes' slope. 
Close to the $70\%$ of computed 
polling stations an increase in the percentage of votes is evident. No matter that in both 
elections this party did not obtain  the largest amount of votes, it appears ruling 
in polling stations arriving at the end of the counting process. 
It is a well known fact, due to historical reasons,  that PRI receives
a lot of votes in geographical regions with a high marginalization index(see
for instance \cite{Pliego}), such
regions are expected to have a slow electoral data processing and transmission to capture centers. 
This might explain why PRI is a sprinter.
In \cite{HernandezSaldanaE4}  the prediction was: 

\bigskip
{\it In the graph of percentage of vote against percentage of processed certificates 
the PRI will change its rate of growth around the time when $70\%$ of the computed certificates arrive. i.e.
this political party has a good final sprint.}
\bigskip

\begin{figure}[ht]
\begin{center}
\includegraphics[width=0.9\columnwidth,angle=0]{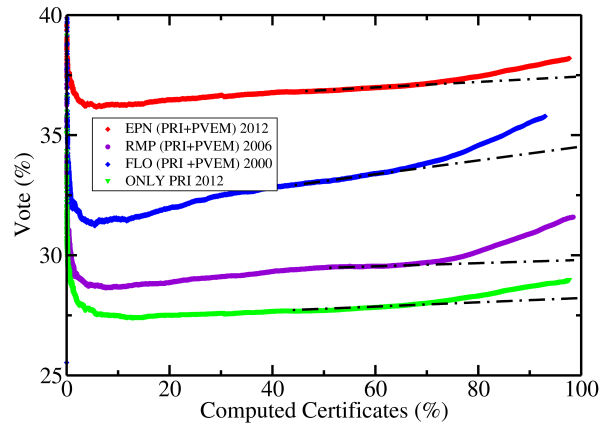}
\end{center}
\caption{{\bf Percentage of votes obtained by the Partido Revolucionario 
Institucional, PRI, presidential candidate in the federal elections of 2012 (upper red line),  2006 (second
from the bottom in violet line), 2000 (third from the bottom in blue line) and percentage of 
votes for PRI alone in July 2012 process (first line in green).} Straight dotted lines are 
included to guide the eye. In 2000 and 2006 only the 
vote for coalitions was admitted. For the 2012 process see explanation in the text. 
Notice that {\it in all} the cases the graphs present a change in slope around $70\%$ of computed
certificates.
}
\label{fig:1}
\end{figure}

In order to test this prediction we report, in Figure \ref{fig:1}, the percentage of vote obtained by PRI against the 
percentage of computed certificates of the polling stations. We report the presidential 
candidates in  2012 (EPN, Enrique Pe\~na Nieto), 2006, (RMP, Roberto Madrazo Pintado) and, 
2000 (FLO, Francisco Labastida Ochoa). For the July 2012 election the rules changed, candidates 
in coalitions appeared in the ballot in coalition and as candidate of 
one party. So, we can differentiate the votes for PRI only, from those obtained by the
options PRI+PVEM and PVEM alone. Hence, we present the case for PRI alone as well. As can be observed in the 
Figure \ref{fig:1}, for all of these cases the PRI changes its growth slope, increasing, in a noticeable way, 
the percentage of votes. No matter if we analyze  PRI alone or the coalition. 
The change in slope is different in all the cases. The 
small party in a coalition presents a typical small party behaviour, (not shown).
Hence, 
{\it From Figure \ref{fig:1} it is clear that prediction ii) has been verified.}

Some small details about Figure \ref{fig:1}. All the polling stations certificates were 
considered in the figure, hence it has small fluctuations that are not appreciable due to 
the plotting character size. The present figure was processed in order to keep the 
file size small. The PREP record ends at a certain hour, usually 24-26 hours after the 
beginning of capture and does not include $100\%$ of the polling stations. So the end of 
records is different for each process.

\subsection*{Prediction iii) The PRI has a smooth vote distribution}

Beyond the important discussion about universal features in vote distribution 
in world wide elections, a corporate political party has been 
extremely regular: the Partido Revolucionario Institucional (PRI). In all the 
previously performed analysis,\cite{HernandezSaldanaE0,HernandezSaldanaE1,HernandezSaldanaE2,HernandezSaldanaE3} 
its vote distribution is a smooth function. 
In reference \cite{HernandezSaldanaE1}, the smooth behaviour of 
this party in federal elections 2000, 2003 and 2006 using the definitive
dataset of Count by District was reported. A similar behaviour has been observed in the
1997 and the 1994 elections by the author but the results remain unpublished. 

The distribution of votes is the histogram of the number of polling 
stations with a certain amount of votes, properly scaled and normalized
in order to have a normalized to unity distribution with a unit mean as well. 
In order to do the comparison with probability 
distribution the amount of votes is ``unfolded'' or ``deconvoluted'' by using 
the average of the number of votes, which scale properly the variable. The resulting
histogram must be normalized to area one.
(In reference \cite{HernandezSaldanaE2} there is
an explanation, but this procedure is standard in data treatment.) 

After this process, fitting a model is possible.
Daisy functions\cite{HernandezSaldana1999}, of different
ranks, were tested with success  for the 2000, 2003 and 2006 electoral processes for  
president and for both chambers.  The only free parameter in this  model is the rank, $r$, and is written as:
\begin{equation}\label{DaisyDist}
P_{r} (s) = \frac{(r+1)^{r+1}}{\Gamma(r+1)} s^r \exp[-(r+1)s].
\end{equation}
With $r$ an integer and $\Gamma(\cdot)$ the Gamma function.

However, this distribution  is a particular 
case of a more general distribution named the Gamma distribution. It is characterized by 
two real free parameters, $\alpha$ and $\theta$ 
\cite{GammaDistributionCRC,GammaDistributionMath} and written as: 
\begin{equation}\label{GammaDist}
P_{\Gamma} (s) = \frac{s^{\alpha-1}}{\Gamma(\alpha) \theta ^{\alpha}} 
   \exp[-\frac{s}{\theta}].
\end{equation}
Here the free parameters are real numbers. 
When $\theta = 1/(r+1)$ and $\alpha = r + 1 $ we recover Eq. (\ref{DaisyDist}).

In this term, the third prediction was presented as:

\bigskip
{\it The distribution of votes for PRI, in presidential and both chambers elections,
fit smooth distributions, 
in general a Gamma distribution or Daisy models.}
\bigskip

\begin{figure}[ht]
\begin{center}
\includegraphics[width=0.9\columnwidth,angle=0]{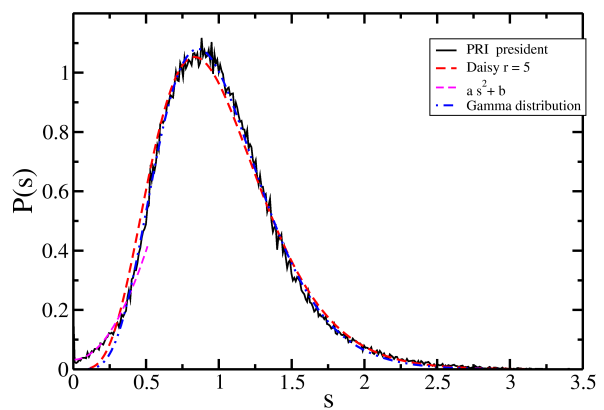}
\end{center}
\caption{{\bf Comparison of the July 2012 PRI vote 
distribution for president (black line) with the two models.} 
In broken red line a Daisy model of rank $r=5$, equation (\ref{DaisyDist},) in blue dot-line the 
Gamma distribution, equation (\ref{GammaDist}), with the best fit. And in magenta broken line 
a quadratic polynomial which fits the distribution's beginning. 
See text for details.}
\label{fig:2}
\end{figure}

The result for the 2012 case is presented in Figure \ref{fig:2} and corresponds only to the 
presidential case for the votes for PRI alone. 
We left the other cases for a future work.  There, the normalized histogram is presented
in a black line. It is noticeable that the beginning of the distribution is not compatible 
with the fast decay at the tail and presents an abrupt change in slope (not shown). 
Such behaviour certainly can be analyzed with 
the Gamma distribution, but we keep the analysis apart
since this kind of change in the slope has been reported for the 2003 intermediate elections.
There, the behaviour corresponds to a different dynamic. 
The beginning of the distribution in Figure \ref{fig:2} is fitted by a quadratic polynomial with no linear term.  

\begin{figure}[ht]
\begin{center}
\includegraphics[width=0.9\columnwidth,angle=0]{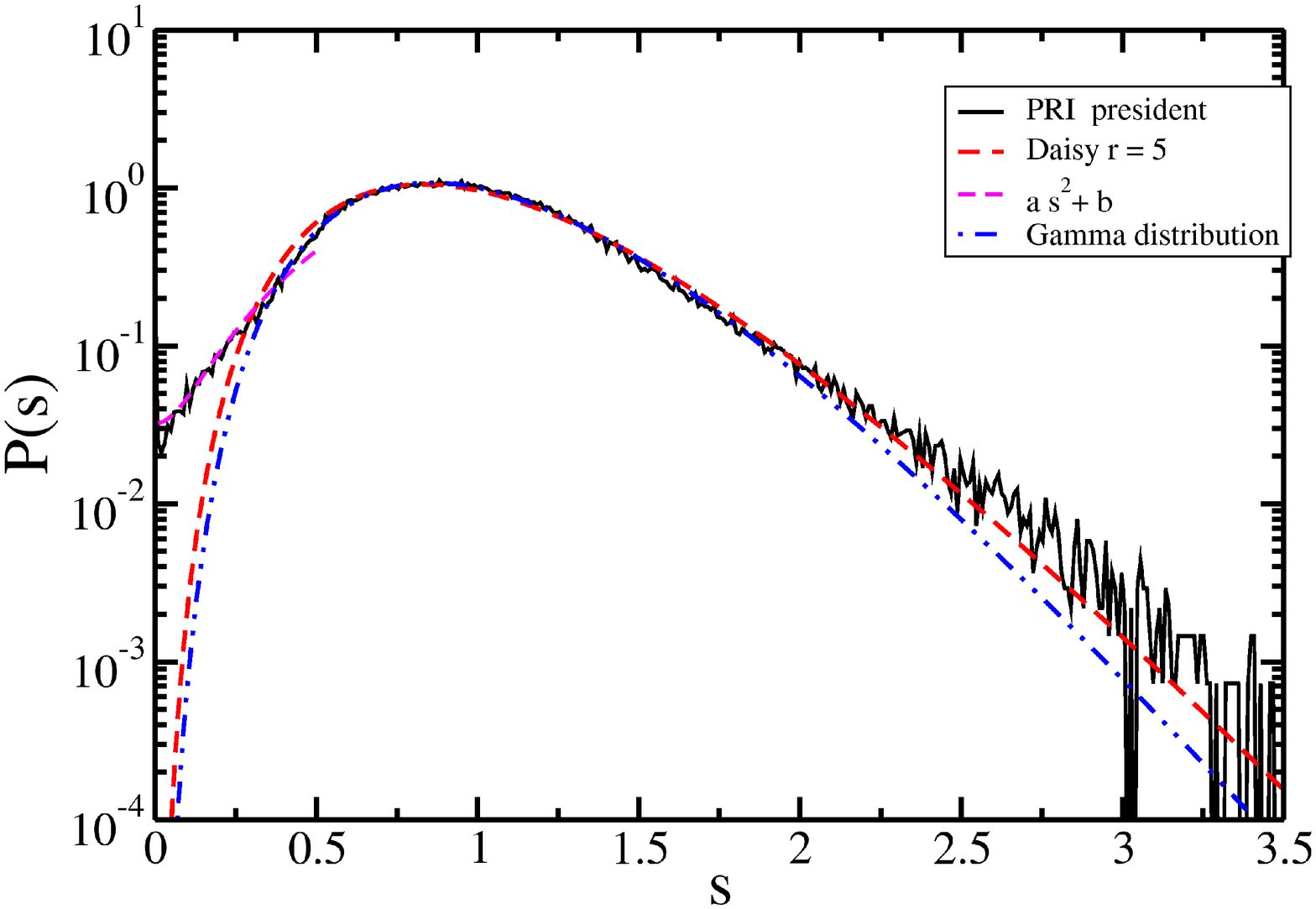}
\end{center}
\caption{{\bf Same as the previous figure
but in log-linear scale.} 
See text for details.}
\label{fig:3}
\end{figure}

For the distribution remaining part we test our two models. In broken red line appears a
Daisy model of rank $r=5$ which follows the curve nicely. The tail is clearly compatible 
with this Daisy as can be seen in Figure \ref{fig:3}, where the same plot is presented but 
in log-linear scale in order to observe the exponential decay. Notice that the Daisy model 
runs inside the range of fluctuations.
Other ranks of Daisy models do not fit
the actual data. 

In order to test how good the Daisy model is, we contrast it with  the  Gamma distribution
with two free parameters, equation (\ref{GammaDist}). The fit was obtained for different starting points, since 
the change in slope is at $s \approx 0.4$. For fittings starting beyond this point the
results are around $\alpha - 1 = 5.8$ and $1/\theta = 6.7$. All the results are compatible 
with a Daisy model, since the relation between $\alpha$ and $\theta$ remains as
$1/\theta \approx \alpha $.
Hence, {\it prediction  iii) was fulfilled, PRI has a smooth vote distribution described
by Daisy models.}

A detailed analysis of the exact values of the parameters is irrelevant at this moment 
since we do not have a theoretical model that explains this smooth behaviour. There are efforts
in this way \cite{HernandezSaldanaE2,HernandezSaldanaE3}. 
Additionally, the results can be a mixture of dynamics since the histogram is built up using  the complete 
database and does not consider differences in district or state, or if the state has been ruled by  PRI for 
a long time. It is important to know
that in several Mexican states PRI rules since 1929. The analysis of PRI's distribution of 
votes performed by state and district is in progress.

\section*{ Conclusions }

Any scientific work must provide predictions, even when they could be based 
only on  empirical observations. To have a valid theoretical framework for the regularities  is 
a much more satisfactory result, but social systems are 
not well understood and this opens wide opportunities for research and for new
multidisciplinary approaches. 

In this paper I offer evidence that Mexican elections, as many others in several countries
and years, present 
regularities. 
The first fulfilled prediction is of a general nature and it says that 
errors are always present, due to honest mistakes or to intentional cheating. 
Here we show that for the third time in a row, mexican elections show 
characteristic distributions of error in the self-consistency records. 
The reason for this behavior is unclear and deserves a separate analysis 
\cite{HernandezSaldana_IIbarra} since they appeared in datasets that correspond
to contrasting political environments: 1)
lack of suspicion of fraud in presidential election (July 2000) with the defeat 
of the long time ruling party, 2) large suspicion of fraud (July 2006) and, 3) 
comeback to power of the ancient political party. Hence, this kind of behavior 
appeared in all of them. Certainly, local fraud by {\it all  political parties} in 
Mexico is well documented, and we hope that some of the common practices appear under 
a much more detailed analysis. 

The second succesful prediction is a result of
history and geography: PRI has a  well established promotional system that has ruled for 
many years. So, no magic intelligence needs to be invoked in order 
to explain this time domain behaviour. Wider studies could confirm this.
The third accomplished prediction is a much more
delicate question. The appearance of probability distributions in a process
is, in general, evidence of some sort of general principle behind it. Such is 
the case of Gaussian distributions or power laws. The appearance of Daisy models 
in all of the PRI electoral process could open a door to understand corporative practices of 
parties around the world. 

\section{Discussion}

\section{Materials and Methods}
All the used datasets are public via the web page of IFE \cite{IFEweb}. Some
dataset are available upon request through the IFE authorities \cite{infomex_ife}
or the author's e-mail. The files in text format contain empty 
fields or comments, some of them are explicit in the annexed documentation. 
For simplicity I did not consider any polling station recorded with an empty
field. The analysis of the error distribution of 2006 PREP was performed 
again with the data base of {\it accepted} polling station certificates.  
All the data treament was made in fortran 77 and the source code 
is available from the author.
\section*{Acknowledgments}
This work was supported by PROMEP/SEP and CONACyT. 
I thank to E. V\'arguez for encouragement and 
support. The author  acknowledges the hospitality of the  
Auberge V\'arguez-Villanueva where this work was ended.


\end{document}